\documentclass[conference,letterpaper]{IEEEtran}
\usepackage[utf8]{inputenc}

\usepackage[T1]{fontenc}
\usepackage{color}
\usepackage{amsmath}
\usepackage{cite}
\usepackage{array}
\usepackage{stfloats}
\usepackage{url}
\usepackage{amsmath,amssymb}
\usepackage{epsfig,fancyhdr}
\usepackage{CJKutf8}
\usepackage{times}
\usepackage{graphicx}
\usepackage{multirow}
\usepackage{stfloats}
\usepackage{balance}

\usepackage{bm}

\newtheorem{thm}{Theorem}

\newtheorem{defin}{Definition}
\newtheorem{lemma}{Lemma}

\newtheorem{eg}{Example}
\newtheorem{remark}{Remark}

\begin{document}
%
\title{Designing Two-Dimensional Complete Complementary Codes for Omnidirectional Transmission in Massive MIMO Systems}
\author{
\IEEEauthorblockN{Cheng-Yu~Pai\begin{CJK}{UTF8}{bsmi} (白承祐)\end{CJK}$^{\scriptscriptstyle \dag}$,~Zilong~Liu\begin{CJK}{UTF8}{bsmi} (劉子龍)\end{CJK}$^{\scriptscriptstyle \ddag}$,~You-Qi~Zhao\begin{CJK}{UTF8}{bsmi} (趙宥齊)\end{CJK}$^{\scriptscriptstyle \dag}$,\\
~Zhen-Ming~Huang\begin{CJK}{UTF8}{bsmi} (黃振銘)\end{CJK}$^{\scriptscriptstyle \dag}$,~and Chao-Yu~Chen\begin{CJK}{UTF8}{bsmi} (陳昭羽)\end{CJK}$^{\scriptscriptstyle \dag}$}
\IEEEauthorblockA{$^{\scriptscriptstyle \dag}$Department of Engineering Science, National Cheng Kung University, Tainan 701, Taiwan, R.O.C.\\
$^{\scriptscriptstyle \ddag}$School of Computer Science and Electronic Engineering, University of Essex, UK\\
Email: $^{\scriptscriptstyle \dag}$\{n98081505, n96091091, n98101012, super\}@mail.ncku.edu.tw, $^{\scriptscriptstyle \ddag}$zilong.liu@essex.ac.uk}
}


\maketitle
\footnotetext[1]{This work was supported in part by the Ministry of Science and Technology,
Taiwan, R.O.C., under Grants MOST 109--2628--E--006--008--MY3 and MOST 110--2224--E--305--001.}

\begin{abstract}
  This paper presents an efficient construction of two-dimensional (2D) complete complementary codes (CCCs) for their modern application as omnidirectional precoding matrices in massive MIMO systems to attain enhanced cell coverage. Unlike the traditional 1D CCCs, little progress has been made on efficient and systematic constructions of the 2D counterpart. In contrast to the existing recursive constructions with the aid of various sequence operations, certain 1D seed sequences or 2D arrays, we propose to use 2D generalized Boolean functions for direct synthesis of 2D CCCs. Simulation results show that the proposed 2D CCCs appear to be good candidates for precoding matrices to achieve omnidirectional transmission in massive MIMO systems.
\end{abstract}
\begin{IEEEkeywords}
Generalized Boolean Function, Complete Complementary Code (CCC), Omnidirectional MIMO Transmission, Precoding Matrices, Uniform Rectangular Arrays (URAs).
\end{IEEEkeywords}
\section{Introduction}
The study of complementary codes (CCs) dates back to the early 1950s in Golay's celebrated publications on a class of sequence pairs (with impulse-like aperiodic autocorrelation sums) for the design of infrared multislit spectrometry \cite{Golay1951,Golay1961}. Over the years, this topic has been gaining increasing research attention as diverse applications of CCs in engineering especially in wireless communications \cite{Golay_RM,Chen_CDMA,Liu_2014,Liu14,CS_sync,Wang2007} are found. Due to the limitation on the set sizes of CCs, the CCs were extended to the Z-complementary sets (ZCSs) to have larger set sizes \cite{ZCP-1st,Wu_18,Sarkar_19}. In the literature, extensive research works are concerned with CCs/ZCSs whose correlations are 1D functions of the time-shifts only \cite{Tseng72,N_shift,Chen08,Wu_20, Pai_20_Electron, Tian_212}. By contrast, this paper focuses efficient design of 2D complete complementary codes (CCCs) which enjoy the maximum set size as well as ideal aperiodic auto-correlations and cross-correlation properties in 2D domain.

In recent years, a major driving force of 2D CCCs is for omnidirectional transmission in massive multi-input and multi-output (MIMO) systems \cite{Meng2016,Lu2020}. Such an application is desired when a base-station (BS) intends to broadcast certain common messages to all the randomly distributed user equipments (UEs) within the current cell, including those at the cell edges, through a public signaling channel. A promising approach is to send through a stream of space-time block coded (STBC) symbols from one or more uniform rectangular arrays (URAs) of the BS such that all the UEs can receive identical signal power for any angle of incidence. It has been shown that such an optimal omnidirectional transmission can be attained when 2D CCCs are adopted as precoding matrices in the URAs \cite{Jiang_19,Li_19,Su2019,Li2021}. Other applications of 2D CCCs include measuring and identification in higher dimensional systems \cite{Luke_85}, image change and motion detection \cite{Khamy2004}, 2D multi-carrier code-division multiple access \cite{2DCDMA_2004,2D_QCCC_CDMA}.

In most of the current state-of-the-art, 2D CCCs are obtained by recursively applying various sequence operations to certain seed 1D sequences or 2D arrays \cite{Luke_85,2D_CCC,Lu2020,Jiang_19,Li_19,Su2019,Li2021}. In \cite{Pai_20,Pai_22}, we presented constructions of 2D mutually orthogonal Golay complementary array pairs (GCAP) which are essentially 2D CCCs with the set size of two only. Recently,\cite{Das_20} constructed 2D Z-complementary array sets (ZCASs) through recursive product of certain matrices of generating polynomials, which may include the 2D CCCs as a special case when the set size equals the number of constituent arrays and the zero-correlation zone covers the entire 2D shift domain. Very recently, \cite{Wang_2021} presented a method to construct higher-dimensional CCCs based on para-unitary (PU) matrices and explicit Boolean functions can be extracted from the seed PU matrix. Then, 2D CCCs can be obtained by applying the projection mapping method in \cite{GCAP_Fiedler_08}. Besides, the design of 2D Z-complementary array pair (ZCAP) has also attracted increasing research attention \cite{Pai_21, Roy_20, Pai_212}.

Direct constructions of 2D CCCs can lead to reduced hardware storage and faster code generation. Motivated by the above background, we propose to use 2D generalized Boolean functions (GBFs) for efficient synthesis of 2D CCCs whose array parameters (e.g., array size, set size, number of constituent arrays) take any power-of-two integer values. {This settles our open problem raised in \cite{Pai_20} on direct constructions of 2D CCCs. Moreover, the proposed 2D CCCs include those from \cite[Th. 6 and Th. 7]{Pai_20} as a special case.} Furthermore, simulation results validate that our proposed 2D CCCs are good candidates for precoding matrices to achieve optimal omnidirectional transmission.

The rest of this paper is outlined as follows. In Section \ref{sec:background}, some preliminaries and notations are given. The concept of 2D generalized Boolean functions is introduced in Section \ref{sec:2Dboolean}. A new construction of 2D CCCs is presented in Section \ref{sec:CCC}. The simulation results are provided in Section \ref{sec:simulation}. Finally, we conclude this work in Section \ref{sec:conclusion}.
\section{Preliminaries and Notations}\label{sec:background}
The following notations will be used throughout this paper:
\begin{itemize}
\item $(\cdot)^*$ denotes the complex conjugation.
\item $\mathbb{Z}_q=\{0,1,\ldots,q-1\}$ is the ring of integers modulo $q$.
\item Let $\xi=e^{2\pi \sqrt{-1}/q}$.
\item $q$ is an even integer.
\end{itemize}
\subsection{Definition of 2D Complete Complementary Code (2D CCC)}
Let $\bm C=(C_{g,i})$ be a $q$-ary array of size $L_1 \times L_2$ where $0\leq g< L_1,~0\leq i< L_2$.
\begin{defin}
The 2-D aperiodic cross-correlation function of arrays $\bm C$ and $\bm D$ at shift $(u_1,u_2)$ is defined as
\begin{equation} \label{eq:cross}
\begin{aligned}
&\rho({\bm C},{\bm D};u_1,u_2)=\\
&\begin{cases}
  \sum\limits_{g=0}^{L_1-1-u_1}\sum\limits_{i=0}^{L_2-1-u_2}\xi^{D_{g+u_1,i+u_2}-C_{g,i}}, 0\leq u_1 <L_1,\\ \quad\quad\qquad\qquad\qquad\qquad\qquad\qquad~ 0\leq u_2 <L_2;\\
  \sum\limits_{g=0}^{L_1-1-u_1}\sum\limits_{i=0}^{L_2-1+u_2}\xi^{D_{g+u_1,i}-C_{g,i-u_2}},  0 \leq  u_1 <L_1,\\ \quad\quad\qquad\qquad\qquad\qquad\qquad\qquad -L_2 < u_2 < 0;\\
  \sum\limits_{g=0}^{L_1-1+u_1}\sum\limits_{i=0}^{L_2-1+u_2}\xi^{D_{g,i}-C_{g-u_1,i-u_2}},  -L_1< u_1 < 0,\\ \quad\quad\qquad\qquad\qquad\qquad\qquad\qquad~ -L_2< u_2 < 0;\\
  \sum\limits_{g=0}^{L_1-1+u_1}\sum\limits_{i=0}^{L_2-1-u_2}\xi^{D_{g,i+u_2}-C_{g-u_1,i}},  -L_1 < u_1 <0,\\ \quad\quad\qquad\qquad\qquad\qquad\qquad\quad\qquad~ 0 \leq u_2 <L_2.
\end{cases}
\end{aligned}
\end{equation}
If ${\bm C}={\bm D}$, $\rho({\bm C},{\bm C};u_1,u_2)$ is called the 2-D aperiodic autocorrelation function of ${\bm C}$ and referred to as $\rho({\bm C};u_1,u_2)$. Note that $\rho({\bm C};u_1,-u_2)=\rho^{*}({\bm C};-u_1,u_2)$.
\end{defin}
\begin{defin}\label{defin:2DCCC}
For a set of $M$ sets of arrays $G=\{G^{p}|p=0,1,\ldots,M-1\}$, each set $G^{p}=\{{\bm C}^{p}_{0},{\bm C}^{p}_{1},\ldots,{\bm C}^{p}_{N-1}\}$ is composed of $N$ arrays and the array size is $L_1\times L_2$. Let
\begin{equation}
\begin{aligned}
&\rho(G^{p},G^{p'};u_1,u_2)=\sum_{t=0}^{N-1}\rho({\bm C}^{p}_t,{\bm C}^{p'}_t;u_1,u_2)\\
&=
\begin{cases}
  NL_1L_2,~~~(u_1,u_2)=(0,0),~ p=p';\\
  0,~~~(u_1,u_2)\neq(0,0),~ p=p';\\
  0,~~~~ p\neq p'
\end{cases}
\end{aligned}
\end{equation}
where ${\bm C}^{p}_t=(C^{p}_t)_{g,i}$ for $0\leq g< L_1,~0\leq i< L_2$. When $M=N$, the set $G$ is called a 2D CCC, denoted by 2D $(M,N,L_1,L_2)$-CCC. Actually, each set $G^p$ forms a 2D {\em Golay Complementary Array Set} (GCAS), referred to as $(N,L_1,L_2)$-GCAS. Note that a 2D CCC can be reduced to a conventional 1-D $(M,N,L_2)$-CCC by taking $L_1=1$. {Likewise, a 1D $(M,N,L_2)$-CCC comprises $M$ distinct 1D {\em Golay complementary sets} (GCSs), where each GCS is denoted by $(N,L_2)$-GCS.}
\end{defin}
\subsection{2D Generalized Boolean Functions (2D GBFs)}\label{sec:2Dboolean}
In this section, we will introduce a concept of a 2D GBFs \cite{Pai_20}. A 2D GBF is composed of $n+m$ variables $y_1,y_2,\ldots,y_n,x_1,x_2,\ldots,x_m$, mapping from $\mathbb{Z}_2^{n+m}$ to $\mathbb{Z}_q$, where $y_s,x_l \in \{0,1\}$ for $s=1,2,\ldots,n$ and $l=1,2,\ldots,m$. We denote a product of $r$ variables by the monomial of degree $r$. For instance, $x_1x_2y_1y_2$ is a monomial of degree 4. The array $\bm f$ linked with the GBF $f$ of $n+m$ variables can be written as
\begin{equation}
{\bm f}=
\begin{pmatrix}
f_{0,0}& f_{0,1} & \cdots & f_{0,2^{m}-1}\\
f_{1,0} & f_{1,1} & \cdots & f_{1,2^{m}-1}\\
\vdots & \vdots & \ddots & \vdots   \\
f_{2^{n}-1,0} & f_{2^{n}-1,1} & \cdots & f_{2^{n}-1,2^{m}-1}
\end{pmatrix}\end{equation}
by letting $f_{g,i}=f((g_1,g_2,\ldots,g_n),(i_1,i_2,\ldots,i_m))$, where $(g_1,g_2,\ldots,g_n)$ and $(i_1,i_2,\ldots,i_m)$ are binary representation vectors of integers $g=\sum_{h=1}^{n}g_{h}2^{h-1}$ and $i=\sum_{j=1}^{m}i_{j}2^{j-1}$, respectively. For example, if $q=2$, $n=2$, and $m=3$, the associated array of the GBF $f=x_1x_3+x_2y_1+y_2$ is
\begin{equation}{\bm f}=
\begin{pmatrix}
0& 0 & 0 & 0 & 0& 1 & 0 & 1\\
0& 0 & 1 & 1 & 0& 1 & 1 & 0\\
1& 1 & 1 & 1 & 1& 0 & 1 & 0\\
1& 1 & 0 & 0 & 1& 0 & 0 & 1
\end{pmatrix}.\end{equation}
\subsection{Omnidirectional Precoding based on 2D Arrays}\label{sec:Omnidirectional}
We briefly introduce the application of the 2D CCC to omnidirectional transmission in massive MIMO systems with a URA.
For the system model, we consider a downlink transmission from a BS equipped with a URA to single-antenna UEs within a cell\cite{Su2019,Li2021}. Assume that the URA comprises $L_1L_2$ antennas with $L_1$ rows and $L_2$ columns. Let ${\bm A}(\varphi,\theta)$ be the steering matrix of size $L_1 \times L_2$ at the direction $(\varphi,\theta)$ and ${\bm A}(\varphi,\theta)=(A(\varphi,\theta))_{g,i}$ for $g=0,1,\ldots,L_{1}-1,~ i=0,1,\ldots,L_{2}-1$. From \cite{Li2021}, we have
\begin{equation}\label{eq:steering_matrix}
\begin{aligned}
(A(\varphi,\theta))_{g,i}=&e^{-j\frac{2\pi}{\lambda}gd_{y}\sin\varphi\sin\theta-j\frac{2\pi}{\lambda}id_{x}\sin\varphi\cos\theta}\\
&\varphi \in[0,\pi/2],~\theta\in[0,2\pi],
\end{aligned}
\end{equation}
where $\lambda$ is the wavelength of the carrier, $d_x$ and $d_y$ represent the inter-elements along the vertical and horizontal directions of the URA, respectively.
In this paper, we consider an orthogonal STBC $\bm S$ of size $N\times N$. Denote the $(n,t)$-th entry of $\bm S$ by $s_n(t)$ which is beamformed by the precoding matrix ${\bm W}_n$ of size $L_1\times L_2$. Therefore, the transmitted signal mapped onto the $L_1L_2$ antennas of the URA is given by
\begin{equation}
\begin{aligned}
\sum_{n=0}^{N-1} {\bm W}_{n}\cdot s_{n}(t),~t=0,1,{\ldots },N-1.
\end{aligned}
\end{equation}
In the line-of-sight (LOS) channel without multipaths, the received signal for a user at direction $(\varphi,\theta)$ is
\begin{equation}
\begin{aligned}
\sum _{n=0}^{N-1} [{\text{vec}}({\bm A}(\varphi,\theta))^T {\text{vec}}({\bm W}_n)] \cdot s_n(t) + w(t), t=0,{\ldots },N-1
\end{aligned}
\end{equation}
where $\text{vec}(\cdot)$ denotes the vectorization of the matrix by vertically stacking the columns into a column vector and $w(t)$ is the complex additive white Gaussian noise (AWGN).
\begin{lemma}\label{lemma:GCAS OP criteria}
 \cite{Li2021} If the precoding matrices ${\bm W}_0,{\bm W}_1,\ldots,$ ${\bm W}_{N-1}$ of size $L_1\times L_2$ are constituent arrays of a 2D $(N,L_1,L_2)$-GCAS, then the received power $\sum_{n=0}^{N-1}\left|[\text{vec}({\bm A}(\varphi,\theta))^T\text{vec}({\bm W}_n)]\right|^2$ is independent of the direction $(\varphi,\theta)$.
\end{lemma}

\textit{Lemma \ref{lemma:GCAS OP criteria}} states that 2D GCASs can be utilized as the precoding matrices to attain omnidirectional transmission. Note that the 2D CCC can be regarded as a collection of 2D GCASs as mentioned in \textit{Definition \ref{defin:2DCCC}}.
\section{Proposed Direct Construction of 2D CCCs}\label{sec:CCC}
Motivated by the construction of 1D CCCs in \cite{Chen08}, we propose a direct construction of  2D CCCs based on 2D GBFs.
\begin{thm}\label{thm:CCC}
For positive integers $m,n,$ and $k$ with $m,n\geq k$, we let nonempty sets $I_1,I_2,\ldots,I_k$ be a partition of $\{1,2,\ldots,m\}$ and also let another nonempty sets $I^{'}_1,I^{'}_2,\ldots,I^{'}_k$ be a partition of $\{1,2,\ldots,n\}$. We take $m_{\alpha}$ and $n_{\alpha}$ as the orders of $I_{\alpha}$ and $I^{'}_{\alpha}$, respectively. Then we define $\pi_{\alpha}$ as a bijection from $\{1,2,\ldots,m_{\alpha}\}$ to $I_{\alpha}$ and let $\sigma_{\alpha}$ be a bijection from $\{1,2,\ldots,n_{\alpha}\}$ to $I'_{\alpha}$. The generalized Boolean function is
\begin{equation}
\begin{aligned}
f=&\frac{q}{2}\left(\sum_{\alpha=1}^{k}\sum_{l=1}^{m_{\alpha}-1}x_{\pi_{\alpha}(l)}x_{\pi_{\alpha}(l+1)}+\sum_{\alpha=1}^{k}\sum_{l=1}^{n_{\alpha}-1}y_{\sigma_{\alpha}(l)}y_{\sigma_{\alpha}(l+1)}\right)\\
&+\frac{q}{2}\sum_{\alpha=1}^{k}x_{\pi_{\alpha}(m_{\alpha})}y_{\sigma_{\alpha}(1)}+\sum_{l=1}^{m}d_{l}x_l+\sum_{l=1}^{n}w_{l}y_l+w_0
\end{aligned}
\end{equation}
where $d_l,w_l\in \mathbb{Z}_q$. If we let
\begin{equation}\label{eq:CCCset}
\begin{aligned}
G^p=\left\{\bm f+\frac{q}{2}\sum_{\alpha=1}^{k}t_{\alpha}{\bm x}_{\pi_{\alpha}(1)}+\frac{q}{2}\sum_{\alpha=1}^{k}p_{\alpha}{\bm y}_{\sigma_{\alpha}(n_{\alpha})}:t_{\alpha}\in \{0,1\}\right\},
\end{aligned}
\end{equation}
where $p_{\alpha}\in \{0,1\}$ and $p=\sum_{l=1}^{k}p_{l}2^{l-1}$. Then the set $\{G^0,G^1,\ldots,G^{2^{k}-1}\}$ forms a $(2^k,2^k,2^n,2^m)$-CCC.
\end{thm}
\begin{IEEEproof}
The proof is given in Appendix.
\end{IEEEproof}

\begin{table*}[ht]
\tiny
\centering
\caption{The sets $G_0$ and $G_1$ in Example \ref{eg:CCC}} \label{table1}
\begin{tabular}{|l|l|l|l|l|l|}
\hline
\multicolumn{4}{|c|}{\normalsize $G^0$}                          \\[0.15 cm] \hline
\multicolumn{2}{|l|}{$
{\bm C}^{0}_0=
\left( {\begin{array}{cccccccccccccccc}
0&	0&	0&	0&	0&	1&	0&	1&	0&	0&	1&	1&	0&	1&	1&	0\\
0&	0&	0&	0&	1&	0&	1&	0&	0&	0&	1&	1&	1&	0&	0&	1\\
0&	0&	0&	0&	0&	1&	0&	1&	0&	0&	1&	1&	0&	1&	1&	0\\
1&	1&	1&	1&	0&	1&	0&	1&	1&	1&	0&	0&	0&	1&	1&	0\\
0&	0&	0&	0&	0&	1&	0&	1&	1&	1&	0&	0&	1&	0&	0&	1\\
0&	0&	0&	0&	1&	0&	1&	0&	1&	1&	0&	0&	0&	1&	1&	0\\
0&	0&	0&	0&	0&	1&	0&	1&	1&	1&	0&	0&	1&	0&	0&	1\\
1&	1&	1&	1&	0&	1&	0&	1&	0&	0&	1&	1&	1&	0&	0&	1\\
\end{array}}
\right)
$} &  \multicolumn{2}{|l|}{$
{\bm C}^{0}_1=
\left( {\begin{array}{cccccccccccccccc}
0&	1&	0&	1&	0&	0&	0&	0&	0&	1&	1&	0&	0&	0&	1&	1\\
0&	1&	0&	1&	1&	1&	1&	1&	0&	1&	1&	0&	1&	1&	0&	0\\
0&	1&	0&	1&	0&	0&	0&	0&	0&	1&	1&	0&	0&	0&	1&	1\\
1&	0&	1&	0&	0&	0&	0&	0&	1&	0&	0&	1&	0&	0&	1&	1\\
0&	1&	0&	1&	0&	0&	0&	0&	1&	0&	0&	1&	1&	1&	0&	0\\
0&	1&	0&	1&	1&	1&	1&	1&	1&	0&	0&	1&	0&	0&	1&	1\\
0&	1&	0&	1&	0&	0&	0&	0&	1&	0&	0&	1&	1&	1&	0&	0\\
1&	0&	1&	0&	0&	0&	0&	0&	0&	1&	1&	0&	1&	1&	0&	0\\
\end{array}}
\right)
$}\\ \hline
\multicolumn{2}{|l|}{$
{\bm C}^{0}_2=
\left( {\begin{array}{cccccccccccccccc}
0&	0&	1&	1&	0&	1&	1&	0&	0&	0&	0&	0&	0&	1&	0&	1\\
0&	0&	1&	1&	1&	0&	0&	1&	0&	0&	0&	0&	1&	0&	1&	0\\
0&	0&	1&	1&	0&	1&	1&	0&	0&	0&	0&	0&	0&	1&	0&	1\\
1&	1&	0&	0&	0&	1&	1&	0&	1&	1&	1&	1&	0&	1&	0&	1\\
0&	0&	1&	1&	0&	1&	1&	0&	1&	1&	1&	1&	1&	0&	1&	0\\
0&	0&	1&	1&	1&	0&	0&	1&	1&	1&	1&	1&	0&	1&	0&	1\\
0&	0&	1&	1&	0&	1&	1&	0&	1&	1&	1&	1&	1&	0&	1&	0\\
1&	1&	0&	0&	0&	1&	1&	0&	0&	0&	0&	0&	1&	0&	1&	0\\
\end{array}}
\right)
$} & \multicolumn{2}{|l|}{$
{\bm C}^{0}_3=
\left( {\begin{array}{cccccccccccccccc}
0&	1&	1&	0&	0&	0&	1&	1&	0&	1&	0&	1&	0&	0&	0&	0\\
0&	1&	1&	0&	1&	1&	0&	0&	0&	1&	0&	1&	1&	1&	1&	1\\
0&	1&	1&	0&	0&	0&	1&	1&	0&	1&	0&	1&	0&	0&	0&	0\\
1&	0&	0&	1&	0&	0&	1&	1&	1&	0&	1&	0&	0&	0&	0&	0\\
0&	1&	1&	0&	0&	0&	1&	1&	1&	0&	1&	0&	1&	1&	1&	1\\
0&	1&	1&	0&	1&	1&	0&	0&	1&	0&	1&	0&	0&	0&	0&	0\\
0&	1&	1&	0&	0&	0&	1&	1&	1&	0&	1&	0&	1&	1&	1&	1\\
1&	0&	0&	1&	0&	0&	1&	1&	0&	1&	0&	1&	1&	1&	1&	1\\
\end{array}}
\right)
$} \\ \hline
\multicolumn{4}{|c|}{\normalsize $G^1$}                          \\[0.15 cm] \hline
\multicolumn{2}{|l|}{$
{\bm C}^{1}_0=
\left( {\begin{array}{cccccccccccccccc}
0&	0&	0&	0&	0&	1&	0&	1&	0&	0&	1&	1&	0&	1&	1&	0\\
0&	0&	0&	0&	1&	0&	1&	0&	0&	0&	1&	1&	1&	0&	0&	1\\
1&	1&	1&	1&	1&	0&	1&	0&	1&	1&	0&	0&	1&	0&	0&	1\\
0&	0&	0&	0&	1&	0&	1&	0&	0&	0&	1&	1&	1&	0&	0&	1\\
0&	0&	0&	0&	0&	1&	0&	1&	1&	1&	0&	0&	1&	0&	0&	1\\
0&	0&	0&	0&	1&	0&	1&	0&	1&	1&	0&	0&	0&	1&	1&	0\\
1&	1&	1&	1&	1&	0&	1&	0&	0&	0&	1&	1&	0&	1&	1&	0\\
0&	0&	0&	0&	1&	0&	1&	0&	1&	1&	0&	0&	0&	1&	1&	0\\
\end{array}}
\right)
$} &  \multicolumn{2}{|l|}{$
{\bm C}^{1}_1=
\left( {\begin{array}{cccccccccccccccc}
0&	1&	0&	1&	0&	0&	0&	0&	0&	1&	1&	0&	0&	0&	1&	1\\
0&	1&	0&	1&	1&	1&	1&	1&	0&	1&	1&	0&	1&	1&	0&	0\\
1&	0&	1&	0&	1&	1&	1&	1&	1&	0&	0&	1&	1&	1&	0&	0\\
0&	1&	0&	1&	1&	1&	1&	1&	0&	1&	1&	0&	1&	1&	0&	0\\
0&	1&	0&	1&	0&	0&	0&	0&	1&	0&	0&	1&	1&	1&	0&	0\\
0&	1&	0&	1&	1&	1&	1&	1&	1&	0&	0&	1&	0&	0&	1&	1\\
1&	0&	1&	0&	1&	1&	1&	1&	0&	1&	1&	0&	0&	0&	1&	1\\
0&	1&	0&	1&	1&	1&	1&	1&	1&	0&	0&	1&	0&	0&	1&	1\\
\end{array}}
\right)
$}\\ \hline
\multicolumn{2}{|l|}{$
{\bm C}^{1}_2=
\left( {\begin{array}{cccccccccccccccc}
0&	0&	1&	1&	0&	1&	1&	0&	0&	0&	0&	0&	0&	1&	0&	1\\
0&	0&	1&	1&	1&	0&	0&	1&	0&	0&	0&	0&	1&	0&	1&	0\\
1&	1&	0&	0&	1&	0&	0&	1&	1&	1&	1&	1&	1&	0&	1&	0\\
0&	0&	1&	1&	1&	0&	0&	1&	0&	0&	0&	0&	1&	0&	1&	0\\
0&	0&	1&	1&	0&	1&	1&	0&	1&	1&	1&	1&	1&	0&	1&	0\\
0&	0&	1&	1&	1&	0&	0&	1&	1&	1&	1&	1&	0&	1&	0&	1\\
1&	1&	0&	0&	1&	0&	0&	1&	0&	0&	0&	0&	0&	1&	0&	1\\
0&	0&	1&	1&	1&	0&	0&	1&	1&	1&	1&	1&	0&	1&	0&	1\\
\end{array}}
\right)
$} & \multicolumn{2}{|l|}{$
{\bm C}^{1}_3=
\left( {\begin{array}{cccccccccccccccc}
0&	1&	1&	0&	0&	0&	1&	1&	0&	1&	0&	1&	0&	0&	0&	0\\
0&	1&	1&	0&	1&	1&	0&	0&	0&	1&	0&	1&	1&	1&	1&	1\\
1&	0&	0&	1&	1&	1&	0&	0&	1&	0&	1&	0&	1&	1&	1&	1\\
0&	1&	1&	0&	1&	1&	0&	0&	0&	1&	0&	1&	1&	1&	1&	1\\
0&	1&	1&	0&	0&	0&	1&	1&	1&	0&	1&	0&	1&	1&	1&	1\\
0&	1&	1&	0&	1&	1&	0&	0&	1&	0&	1&	0&	0&	0&	0&	0\\
1&	0&	0&	1&	1&	1&	0&	0&	0&	1&	0&	1&	0&	0&	0&	0\\
0&	1&	1&	0&	1&	1&	0&	0&	1&	0&	1&	0&	0&	0&	0&	0\\
\end{array}}
\right)
$} \\ \hline
\end{tabular}
\end{table*}

\begin{table*}[ht]
\caption{Comparisons of 2D $(M,N,L_1,L_2)$-CCCs \label{table2}}
\begin{center}
\begin{tabular}{|c||c|c|c|c|c|c|}\hline
Method                                               & Parameters                & Based on                      & Note                                   \\ \hline
\cite[Th. 3]{Lu2020}                                      &   $(1,N,L_1,L_2)$         &1D $(N,L_1)$-GCSs and 1D $(2,L_2)$-GCSs  &$N,L_1,L_2 \geq 2$        \\ \hline
\cite[Lemma III.3]{Jiang_19}                                      & $(1,N_1N_2,L_1,L_2)$           &1D $(N_1,L_1)$-GCSs and 1D $(N_2,L_2)$-GCSs  &$N_1,N_2,L_1,L_2 \geq 2$     \\ \hline
\cite[Th. III.3]{Li2021}                                      & $(2,2,2L_1,L_2)$           &1D $(2,L_1)$-GCSs and 1D $(2,L_2)$-GCSs  & $L_1,L_2 \geq 2$       \\ \hline
\cite{2D_CCC}                                      & $(MK,MK,{M}^{2},{K}^{2})$           & 1D $(M,M,{M}^{2})$-CCCs and 1D $(K,K,{K}^{2})$-CCCs  & $M,K\geq 2$        \\ \hline
 \cite[Th. 7]{Pai_20}                                      & $(2,2,2^n,2^m)$                & 2D Generalized Boolean functions   & $n,m> 0$        \\ \hline
\cite[Th. 1]{Das_20}                         & $(M,M,M,M)$              & BH matrices of array size $M\times M$      & $M\geq 2$                              \\ \hline
\textit{Theorem \ref{thm:CCC}}~(Proposed)                 & $(2^k,2^k,2^n,2^m)$ & 2D Generalized Boolean functions  & $n,m\geq k > 0$,
\\ \hline
\end{tabular}
\end{center}
 \end{table*}

\begin{remark}\label{rmk:CCC}
If we take $k=1$ in Theorem $\ref{thm:CCC}$, then the proposed 2-D CCCs will be reduced to $(2,2,2^n,2^m)$-CCCs given in \cite[Th. 7]{Pai_20}. Therefore, \cite[Th. 7]{Pai_20} can be viewed as a special case of Theorem \ref{thm:CCC}.
\end{remark}
\begin{eg}\label{eg:CCC}
Taking $q=2$, $m=4$, $n=3$, and $k=2$, we let $I_1=\{1,3\},~I_2=\{2,4\},~I^{\prime}_1=\{1,2\},~I^{\prime}_2=\{3\},~\pi_1(1)=1,~\pi_1(2)=3,~\pi_2(1)=2,~\pi_2(2)=4,~\sigma_1(1)=1,~\sigma_1(2)=2,~\text{and}~\sigma_2(1)=3$. According to Theorem \ref{thm:CCC}, the generalized Boolean function is $f=x_1x_3+x_2x_4+y_1y_2+y_3+x_3y_1+x_4y_3$ by setting $d_l=0$ for $l=1,2,3,4$ and $w_s=0$ for $s=0,1,2,3$. The array sets is given by $G^p=\{\bm f+t_{1}{\bm x}_1+t_2{\bm x}_2+p_1{\bm y}_2+p_2{\bm y}_3:t_1, t_2\in \{0,1\}\}$ for $p=0,1,2,3$. The set $\{G^0,G^1,G^2,G^3\}$ forms a $(4,4,8,16)$-CCC from Theorem \ref{thm:CCC}. We list array sets $G_0=\{{\bm C}^{0}_0,{\bm C}^{0}_1,{\bm C}^{0}_2,{\bm C}^{0}_3\}$ and $G_1=\{{\bm C}^{1}_0,{\bm C}^{1}_1,{\bm C}^{1}_2,{\bm C}^{1}_3\}$, respectively, in Table \ref{table1}. One can verify that $G^0$ and $G^1$ are 2D GCASs, respectively. Also, the cross-correlations of $G^0$ and $G^1$ are zero in every shift.
\end{eg}

In Table \ref{table2}, we compare our proposed construction with existing constructions of 2D CCCs. {Note that the constructions in \cite[Th. 3]{Lu2020} and \cite[Lemma III.3]{Jiang_19} can only generate 2D CCCs with ideal aperiodic autocorrelation sums only. In \cite[Th. III.3]{Li2021}, 2D CCCs can be constructed by existing 1D GCSs where the set size is restricted to two. It can be seen that our proposed 2D CCCs enjoy the most flexible set sizes and array sizes. Also, our proposed construction is direct instead of using any 1D CCCs, 1D GCSs, or 2D BH matrices as kernels.}
\section{Simulation Results}\label{sec:simulation}
\begin{figure}[ht!]
	\centering
	\includegraphics[width = 2.8in]{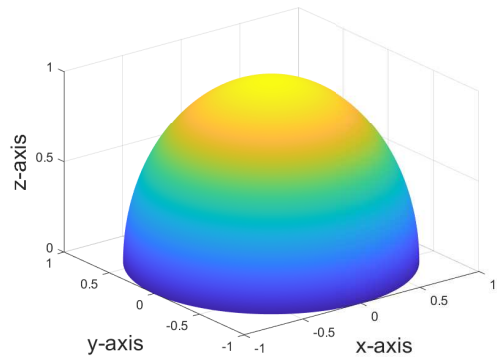}
    \caption{Power radiation pattern generated by the 2D CCC-based scheme for $8\times 16 $ URA.}
    \label{power}
\end{figure}

\begin{figure}[ht!]
	\centering
	\vspace{-0.2cm}\includegraphics[width=2.9in]{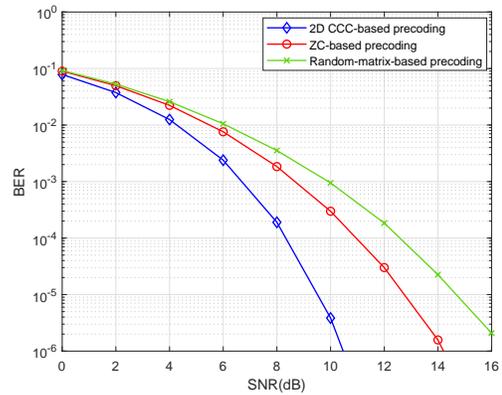}
    \caption{BER performance comparison for different precoding schemes with an $8\times 16$ URA.\label{BER}}
\end{figure}

In this section, we employ our proposed 2D CCCs as precoding matrices and present numerical simulation results. Throughout the simulation, we consider a massive MIMO system (over a LOS channel corrupted by AWGN) with a URA of size $8\times 16$, i.e., $L_1=8$ and $L_2=16$. Let $N=4$ and the $4\times 4$ orthogonal STBC be given by
\begin{equation}{\label{STBC}}
\begin{aligned}
{\bm S}= \left(\begin{matrix}s_0 & -s_1 & -s_2 & -s_3\\ s_1& s_0 & s_3 & -s_2\\ s_2&-s_3&s_0&s_1\\ s_3&s_2&-s_1&s_0 \end{matrix}\right),
\end{aligned}
\end{equation}
where the symbols in $\bm S$ are all modulated by binary phase shift keying (BPSK) \cite{Tarokh_99}. Considering the steering matrix in (\ref{eq:steering_matrix}) and the set $G^{0}=\{{\bm C}^{0}_0,{\bm C}^{0}_1,{\bm C}^{0}_2,{\bm C}^{0}_3\}$ in \textit{Example \ref{eg:CCC}}, let precoding matrices $\{{\bm W}_0,{\bm W}_1,{\bm W}_2,{\bm W}_3\}=\{{\bm C}_0^0,{\bm C}_1^0,{\bm C}_2^0,{\bm C}_3^0\}$. The power radiation pattern, i.e., ${\sum_{n=0}^{3}\left|[\text{vec}({\bm A}(\varphi,\theta))^T\text{vec}({\bm W}_n)]\right|^2}$, is depicted in Fig. \ref{power}, which can achieve omnidirectional transmission by using our proposed 2D CCC as precoding matrices.
Furthermore, in Fig. \ref{BER}, we illustrate the bit error rate (BER) performance within the LOS channel without multipaths. The Zadoff-Chu sequence based (ZC-based) scheme uses the similar method given in \cite[(95)]{Li2021}, which needs four precoding matrices of size $8\times 16$ obtained from two ZC sequences of lengths 8 and 16, respectively. The random-matrix-based scheme utilizes four $8\times 16$ random matrices with values drawn from $\{+1,-1\}$.
It can be observed that the performance of the 2D CCC-based scheme enjoys about 4 dB gain for BER at $10^{-6}$ compared to the ZC-based scheme. Hence, our proposed 2D CCCs are good candidates for omnidirectional precoding matrices in massive MIMO systems.

\section{Conclusion}\label{sec:conclusion}
In this work, driven by their modern application for omnidirectional transmission in massive MIMO systems, we have introduced a novel direct design of 2D CCCs based on 2D GBFs in Theorem \ref{thm:CCC}. The proposed 2D $(2^k,2^k,2^n,2^m)$-CCCs have flexible set sizes and array sizes, which includes our earlier result in \cite[Th. 7]{Pai_20} as a special case. Furthermore, we have illustrated that the proposed 2D CCCs are good omnidirectional precoding matrices in massive MIMO systems.

\vspace{-1.3cm}
\appendix[Proof of Theorem \ref{thm:CCC}]
\begin{IEEEproof}
For the array ${\bm C}^{p}_t$ constructed from Theorem \ref{thm:CCC}, it can be expressed as
\begin{equation}\label{eq:C_Boolean}
\begin{aligned}
{\bm C}^{p}_t=\bm f+\frac{q}{2}\sum_{\alpha=1}^{k}t_{\alpha}{\bm x}_{\pi_{\alpha}(1)}+\frac{q}{2}\sum_{\alpha=1}^{k}p_{\alpha}{\bm y}_{\sigma_{\alpha}(n_{\alpha})},
\end{aligned}
\end{equation}
where
\begin{equation}
{\bm C}^{p}_t=
\begin{pmatrix}
(C^{p}_t)_{0,0}& (C^{p}_t)_{0,1} & \cdots & (C^{p}_t)_{0,2^{m}-1}\\[0.2 cm]
(C^{p}_t)_{1,0} & (C^{p}_t)_{1,1} & \cdots & (C^{p}_t)_{1,2^{m}-1}\\
\vdots & \vdots & \ddots & \vdots   \\
(C^{p}_t)_{2^{n}-1,0} & (C^{p}_t)_{2^{n}-1,1} & \cdots & (C^{p}_t)_{2^{n}-1,2^{m}-1}
\end{pmatrix}\end{equation}
for $0\leq t,p< 2^{k}$.

In the first part, we would like to prove that every $G^{p}=\{{\bm C}^{p}_0,{\bm C}^{p}_1,\ldots,{\bm C}^{p}_{2^{k}-1}\}$ is a 2D GCAS for $p=0,1,\ldots, 2^{k}-1$. Namely,
\begin{equation}\label{eq:auto_c}
\begin{aligned}
&\sum_{t=0}^{2^{k}-1}\sum\limits_{g=0}^{2^{n}-1-u_1}\sum\limits_{i=0}^{2^{m}-1-u_2}\left(\xi^{(C^{p}_{t})_{g+u_1,{i+u_2}}-(C^{p}_{t})_{g,i}}\right)\\
&=\sum\limits_{g=0}^{2^{n}-1-u_1}\sum\limits_{i=0}^{2^{m}-1-u_2}\sum_{t=0}^{2^{k}-1}\left(\xi^{(C^{p}_{t})_{g+u_1,{i+u_2}}-(C^{p}_{t})_{g,i}}\right)=0
\end{aligned}
\end{equation}
for $0\leq u_{1}<2^{n},~0\leq u_{2}<2^m$, and $(u_1,u_2)\neq (0,0)$. Then we let $h=g+u_1,j=i+u_2$ and also let $(g_1,g_2,\ldots,g_n),(h_1,h_2,\ldots,h_n),(i_1,i_2,\ldots,i_m)$, and $(j_1,j_2,\ldots,j_m)$ be the binary representations of $g,h,i$, and $j$, respectively. Next we consider four cases to demonstrate that (\ref{eq:auto_c}) holds.

{\it Case 1:} If $u_1 \geq 0$ and $u_2> 0$, we suppose $i_{\pi_{\alpha}(1)}\neq j_{\pi_{\alpha}(1)}$ for some $\alpha \in \{1,2,\ldots,k\}$. Then for any array ${\bm C}^{p}_t\in G^{p}$, we can find an array ${\bm C}^{p}_{t^{'}}={\bm C}^{p}_t+(q/2){\bm x}_{\pi_{\alpha}(1)}\in G^{p}$ satisfying $(C^{p}_{t})_{h,j}-(C^{p}_{t})_{g,i}-(C^{p}_{t^{'}})_{h,j}+(C^{p}_{t^{'}})_{g,i}\equiv \frac{q}{2} \pmod q.$
Thus, we have
$\sum_{t=0}^{2^{k}-1}\left(\xi^{(C^{p}_{t})_{h,{j}}-(C^{p}_{t})_{g,i}}\right)=0.$

{\it Case 2:} For $u_1 \geq 0$ and $u_2> 0$, we consider $i_{\pi_{\alpha}(1)}= j_{\pi_{\alpha}(1)}$ for all $\alpha \in \{1,2,\ldots,k\}$. Assume $i_{\pi_{\alpha}(\beta)}=j_{\pi_{\alpha}(\beta)}$ for $\alpha=1,2,\ldots,\hat{\alpha}-1$ and $\beta=1,2,\ldots, m_{\alpha}$. Then we set $\hat{\beta}$ as the smallest integer such that $i_{\pi_{\hat{\alpha}}(\hat{\beta})}\neq j_{\pi_{\hat{\alpha}}(\hat{\beta})}$. Let $i'$ and $j'$ be integers distinct from $i$ and $j$, respectively, only in one position. That is, $i^{\prime}_{\pi_{\hat{\alpha}}(\hat{\beta}-1)}=1-i_{\pi_{\hat{\alpha}}(\hat{\beta}-1)}$ and $j^{\prime}_{\pi_{\hat{\alpha}}(\hat{\beta}-1)}=1-j_{\pi_{\hat{\alpha}}(\hat{\beta}-1)}$. Hence, we have
\begin{equation}
\begin{aligned}
(C^{p}_{t})_{g,i'}-(C^{p}_{t})_{g,i}=&\frac{q}{2}\left(i_{\pi_{\hat{\alpha}}(\hat{\beta}-2)}i^{\prime}_{\pi_{\hat{\alpha}}(\hat{\beta}-1)}-i_{\pi_{\hat{\alpha}}(\hat{\beta}-2)}i_{\pi_{\hat{\alpha}}(\hat{\beta}-1)}\right. \\
&\left. +i^{\prime}_{\pi_{\hat{\alpha}}(\hat{\beta}-1)}i_{\pi_{\hat{\alpha}}(\hat{\beta})}-i_{\pi_{\hat{\alpha}}(\hat{\beta}-1)}i_{\pi_{\hat{\alpha}}(\hat{\beta})} \right)\\
&+d_{\pi_{\hat{\alpha}}(\hat{\beta}-1)}i^{\prime}_{\pi_{\hat{\alpha}}(\hat{\beta}-1)}-d_{\pi_{\hat{\alpha}}(\hat{\beta}-1)}i_{\pi_{\hat{\alpha}}(\hat{\beta}-1)}\\
\equiv &\frac{q}{2}(i_{\pi_{\hat{\alpha}}(\hat{\beta}-2)}+i_{\pi_{\hat{\alpha}}(\hat{\beta})})\\
&+d_{\pi_{\hat{\alpha}}(\hat{\beta}-1)}(1-2i_{\pi_{\hat{\alpha}}(\hat{\beta}-1)}) \pmod q.
\end{aligned}
\end{equation}
Since $i_{\pi_{\hat{\alpha}}(\hat{\beta}-1)}=j_{\pi_{\hat{\alpha}}(\hat{\beta}-1)}$ and $i_{\pi_{\hat{\alpha}}(\hat{\beta}-2)}=j_{\pi_{\hat{\alpha}}(\hat{\beta}-2)}$, we have
\begin{equation}
\begin{aligned}
&(C^{p}_{t})_{h,j}-(C^{p}_{t})_{g,i}-(C^{p}_{t})_{h,j'}+(C^{p}_{t})_{h,i'}\\
&\equiv \frac{q}{2}(i_{\pi_{\hat{\alpha}}(\hat{\beta}-2)}-j_{\pi_{\hat{\alpha}}(\hat{\beta}-2)}+i_{\pi_{\hat{\alpha}}(\hat{\beta})}-j_{\pi_{\hat{\alpha}}(\hat{\beta})})\\
&+d_{\pi_{\hat{\alpha}}(\hat{\beta}-1)}(2j_{\pi_{\hat{\alpha}}(\hat{\beta}-1)}-2i_{\pi_{\hat{\alpha}}(\hat{\beta}-1)})\\
&\equiv \frac{q}{2}(i_{\pi_{\hat{\alpha}}(\hat{\beta})}-j_{\pi_{\hat{\alpha}}(\hat{\beta})})\equiv \frac{q}{2} \pmod q.
\end{aligned}
\end{equation}
implying $\sum_{t=0}^{2^{k}-1}\left(\xi^{(C^{p}_{t})_{h,j}-(C^{p}_{t})_{g,i}}+\xi^{(C^{p}_{t})_{h,j'}-(C^{p}_{t})_{g,i'}}\right)=0$.

{\it Case 3:}  For $u_1 > 0$ and $u_2= 0$, we take $g_{\sigma_{{\alpha}}(1)}\neq h_{\sigma_{{\alpha}}(1)}$ for some $\alpha \in \{1,2,\ldots,k\}$. Using the similar logic, we let $i^{\prime}_{\pi_{{\alpha}}(m_{\alpha})}=1-i_{\pi_{{\alpha}}(m_{\alpha})}$. By following the similar arguments as given in Case 2, we can obtain $\sum_{t=0}^{2^{k}-1}\left(\xi^{(C^{p}_{t})_{h,i}-(C^{p}_{t})_{g,i}}+\xi^{(C^{p}_{t})_{h,i'}-(C^{p}_{t})_{g,i'}}\right)=0.$

{\it Case 4:} For $u_1 > 0$ and $u_2= 0$, we consider $g_{\sigma_{{\alpha}}(1)}= h_{\sigma_{{\alpha}}(1)}$ for all $\alpha \in \{1,2,\ldots,k\}$ and $g_{\sigma_{\alpha}(\beta)}=h_{\sigma_{\alpha}(\beta)}$ for $\alpha=1,2,\ldots,\hat{\alpha}-1$ and $\beta=1,2,\ldots,n_{\alpha}$. If $\hat{\beta}$ is the smallest number with $g_{\sigma_{\hat{\alpha}}(\hat{\beta})}\neq h_{\sigma_{\hat{\alpha}}(\hat{\beta})}$, we let $g^{\prime}_{\sigma_{\hat{\alpha}}(\hat{\beta}-1)}=1-g_{\sigma_{\hat{\alpha}}(\hat{\beta}-1)}$ and $h^{\prime}_{\sigma_{\hat{\alpha}}(\hat{\beta}-1)}=1-h_{\sigma_{\hat{\alpha}}(\hat{\beta}-1)}$. Then using the similar argument as mentioned in Case 2, we can obtain $\sum_{t=0}^{2^{k}-1}\left(\xi^{(C^{p}_{t})_{h,i}-(C^{p}_{t})_{g,i}}+\xi^{(C^{p}_{t})_{h',i}-(C^{p}_{t})_{g',i}}\right)=0.$
From Case 1 to Case 4, we prove that (\ref{eq:auto_c}) holds.

In the second part, we will demonstrate that any two distinct array sets $G^s$ and $G^v$ possess ideal cross-correlation values for $0\leq s\neq v \leq 2^{k}-1$. In other words,
 \begin{equation}\label{eq:v&s}
\hspace{-0.08cm}\begin{aligned}
&\sum_{t=0}^{2^{k}-1}\sum\limits_{g=0}^{2^{n}-1-u_1}\sum\limits_{i=0}^{2^{m}-1-u_2}\left(\xi^{(C^{s}_{t})_{g+u_1,{i+u_2}}-(C^{v}_{t})_{g,i}}\right)\\
=&\sum\limits_{g=0}^{2^{n}-1-u_1}\sum\limits_{i=0}^{2^{m}-1-u_2}\sum_{t=0}^{2^{k}-1}\left(\xi^{(C^{s}_{t})_{g+u_1,{i+u_2}}-(C^{v}_{t})_{g,i}}\right)=0.
\end{aligned}
\end{equation}
Similarly, we let $h=g+u_1,~j=i+u_2$, and consider five cases below.

{\it Case 1:} Consider $u_1\geq 0,~u_2>0$ and $i_{\pi_{\alpha}(1)}\neq j_{\pi_{\alpha}(1)}$ for some $\alpha \in \{1,2,\ldots,k\}$. By following a similar derivation in Case 1 of the first part, we can obtain $\sum_{t=0}^{2^{k}-1}\left(\xi^{(C^{s}_{t})_{g+u_1,{i+u_2}}-(C^{v}_{t})_{g,i}}\right)=0.$

{\it Case 2:} Taking $u_1 \geq 0$ and $u_2> 0$, we suppose $i_{\pi_{\alpha}(1)}= j_{\pi_{\alpha}(1)}$ for all $\alpha \in \{1,2,\ldots,k\}$. The similar result can be obtained as mentioned in Case 2 of the first part. Thus, we have $\sum_{t=0}^{2^{k}-1}\left(\xi^{(C^{s}_{t})_{h,j}-(C^{v}_{t})_{g,i}}+\xi^{(C^{s}_{t})_{h,j'}-(C^{v}_{t})_{g,i'}}\right)=0.$

{\it Case 3:} For $u_1 > 0$ and $u_2= 0$, we assume $g_{\sigma_{{\alpha}}(1)}\neq h_{\sigma_{{\alpha}}(1)}$ for some $\alpha \in \{1,2,\ldots,k\}$. We can obtain the similar result $\sum_{t=0}^{2^{k}-1}\left(\xi^{(C^{s}_{t})_{h,i}-(C^{v}_{t})_{g,i}}+\xi^{(C^{s}_{t})_{h,i'}-(C^{v}_{t})_{g,i'}}\right)=0$
 as arguing in Case 3 of the first part.

 {\it Case 4:} Suppose $u_1 > 0,~u_2= 0$, and $g_{\sigma_{{\alpha}}(1)}=h_{\sigma_{{\alpha}}(1)}$ for all $\alpha \in \{1,2,\ldots,k\}$. Following the similar argument provided in Case 4 of the first part, we have $\sum_{t=0}^{2^{k}-1}\left(\xi^{(C^{s}_{t})_{h,i}-(C^{v}_{t})_{g,i}}+\xi^{(C^{s}_{t})_{h',i}-(C^{v}_{t})_{g',i}}\right)=0.$

 {\it Case 5:} Lastly, it remains to show that (\ref{eq:v&s}) holds for $(u_1,u_2)=(0,0)$. From (\ref{eq:C_Boolean}), we can obtain
\begin{equation}\label{eq:noshift}
\begin{aligned}
\hspace{-0.08cm}(C^{s}_{t})_{g,i}-(C^{v}_{t})_{g,i}\equiv\frac{q}{2}\sum_{\alpha=1}^{k}(s_{\alpha}- v_{\alpha})g_{\sigma_{\alpha}(n_{\alpha})} \pmod q
\end{aligned}
\end{equation}
where $s_{\alpha}$, $v_{\alpha}$, and $g_{\sigma_{\alpha}(n_{\alpha})}$ are the $\alpha$-th bit and the ${\sigma_{\alpha}(n_{\alpha})}$-th bit of binary representations of $s$, $v$, and $g$, respectively. We can observe that (\ref{eq:noshift}) is the linear combination of the term $g_{\sigma_{\alpha}(n_{\alpha})}$. Note that $g=\sum_{l=1}^{n}g_{l}2^{l-1}$. For $g=0,1,\ldots,2^{n}-1$, there are $2^{n-1}$ pairs fulfilling $\xi^{(C^{s}_{t})_{g,i}-(C^{v}_{t})_{g,i}}=\xi^{q/2}=-1$ and $2^{n-1}$ pairs satisfying $\xi^{(C^{s}_{t})_{g,i}-(C^{v}_{t})_{g,i}}=\xi^{0}=1$. Therefore, we have $\sum\limits_{i=0}^{2^{m}-1}\sum\limits_{g=0}^{2^{n}-1}\left(\xi^{(C^{s}_{t})_{g,{i}}-(C^{v}_{t})_{g,i}}\right)=0.$

Combining these five cases, we complete the proof.
\end{IEEEproof}

\balance
\bibliographystyle{IEEEtran}
\bibliography{IEEEabrv,ref}

\begin{thebibliography}{10}
\providecommand{\url}[1]{#1}
\csname url@samestyle\endcsname
\providecommand{\newblock}{\relax}
\providecommand{\bibinfo}[2]{#2}
\providecommand{\BIBentrySTDinterwordspacing}{\spaceskip=0pt\relax}
\providecommand{\BIBentryALTinterwordstretchfactor}{4}
\providecommand{\BIBentryALTinterwordspacing}{\spaceskip=\fontdimen2\font plus
\BIBentryALTinterwordstretchfactor\fontdimen3\font minus
  \fontdimen4\font\relax}
\providecommand{\BIBforeignlanguage}[2]{{%
\expandafter\ifx\csname l@#1\endcsname\relax
\typeout{** WARNING: IEEEtran.bst: No hyphenation pattern has been}%
\typeout{** loaded for the language `#1'. Using the pattern for}%
\typeout{** the default language instead.}%
\else
\language=\csname l@#1\endcsname
\fi
#2}}
\providecommand{\BIBdecl}{\relax}
\BIBdecl

\bibitem{Golay1951}
M.~J.~E. Golay, ``Static multislit spectrometry and its application to the
  panoramic display of infrared spectra,'' \emph{J. Opt. Soc. Amer.}, vol.~41,
  no.~7, p. 468–472, 1951.

\bibitem{Golay1961}
------, ``Complementary series,'' \emph{IRE Trans. Inf. Theory}, vol.~7, no.~2,
  pp. 82--87, Apr. 1961.

\bibitem{Golay_RM}
J.~A. Davis and J.~Jedwab, ``Peak-to-mean power control in {OFDM}, {Golay}
  complementary sequences, and {Reed-Muller} codes,'' \emph{{IEEE} Trans. Inf.
  Theory}, vol.~45, no.~7, pp. 2397--2417, Nov. 1999.

\bibitem{Chen_CDMA}
H.-H. Chen, J.-F. Yeh, and N.~Suehiro, ``A multicarrier {CDMA} architecture
  based on orthogonal complete complementary codes for new generations of
  wideband wireless communications,'' \emph{{IEEE} Commun. Mag.}, vol.~39, pp.
  126--134, Oct. 2001.

\bibitem{Liu_2014}
Z.~Liu, Y.~L. Guan, and U.~Parampalli, ``New complete complementary codes for
  peak-to-mean power control in multi-carrier {CDMA},'' \emph{{IEEE} Trans.
  Commun.}, vol.~62, pp. 1105--1113, Mar. 2014.

\bibitem{Liu14}
------, ``A new construction of zero correlation zone sequences from
  generalized {Reed-Muller} codes,'' in \emph{Proc.\ {IEEE} Inf. Theory
  Workshop}, Hobart, Australia, Nov. 2014, pp. 591--595.

\bibitem{CS_sync}
J.~M. Groenewald and B.~T. Maharaj, ``{MIMO} channel synchronization using
  {Golay} complementary pairs,'' in \emph{Proc.\ AFRICON 2007}, Windhoek, South
  Africa, Sep. 2007, pp. 1--5.

\bibitem{Wang2007}
S.~Wang and A.~Abdi, ``{MIMO ISI channel estimation using uncorrelated Golay
  complementary sets of polyphase sequences},'' \emph{IEEE Trans. Veh.
  Technol.}, vol.~56, no.~5, p. 3024–3039, 2007.

\bibitem{ZCP-1st}
P.~Fan, W.~Yuan, and Y.~Tu, ``{Z}-complementary binary sequences,''
  \emph{{IEEE} Signal Process. Lett.}, vol.~14, no.~8, pp. 509--512, Aug. 2007.

\bibitem{Wu_18}
S.-W. Wu and C.-Y. Chen, ``Optimal {Z}-complementary sequence sets with good
  peak-to-average power-ratio property,'' \emph{IEEE Signal Process. Lett.},
  vol.~25, no.~10, pp. 1500--1504, Oct. 2018.

\bibitem{Sarkar_19}
P.~Sarkar, S.~Majhi, and Z.~Liu, ``Optimal {Z}-complementary code set from
  generalized {Reed-Muller} codes,'' \emph{{IEEE} Trans. Commun.}, vol.~67,
  no.~3, pp. 1783--1796, Mar. 2019.

\bibitem{Tseng72}
C.-C. Tseng and C.~L. Liu, ``Complementary sets of sequences,'' \emph{{IEEE}
  Trans. Inf. Theory}, vol. IT-18, pp. 644--652, Sep. 1972.

\bibitem{N_shift}
N.~Suehiro and M.~Hatori, ``{$N$-shift} cross-orthogonal sequences,''
  \emph{{IEEE} Trans. Inf. Theory}, vol.~34, pp. 143--146, Jan. 1988.

\bibitem{Chen08}
C.-Y. Chen, C.-H. Wang, and C.-C. Chao, ``Complete complementary codes and
  generalized {Reed-Muller} codes,'' \emph{{IEEE} Commun. Lett.}, vol.~12, pp.
  849--851, Nov. 2008.

\bibitem{Wu_20}
S.-W. Wu, C.-Y. Chen, and Z.~Liu, ``How to construct mutually orthogonal
  complementary sets with non-power-of-two lengths?'' \emph{{IEEE} Trans. Inf.
  Theory}, vol.~67, no.~6, pp. 3464--3472, Jun. 2021.

\bibitem{Pai_20_Electron}
C.-Y. Pai and C.-Y. Chen, ``Construction of complementary sequence sets based
  on complementary pairs,'' \emph{Electron. Lett.}, vol.~56, no.~18, pp.
  966--968, Sep. 2020.

\bibitem{Tian_212}
L.~Tian, X.~Lu, C.~Xu, and Y.~Li, ``New mutually orthogonal complementary sets
  with non-power-of-two lengths,'' \emph{{IEEE} Signal Process. Lett.},
  vol.~28, Jan. 2021.

\bibitem{Meng2016}
X.~Meng, X.~Gao, and X.-G. Xia, ``Omnidirectional precoding based transmission
  in massive {MIMO} systems,'' \emph{IEEE Trans. Commun.}, vol.~64, no.~1, pp.
  174--186, Jan. 2016.

\bibitem{Lu2020}
A.-A. Lu, X.~Gao, and X.~Meng, ``Omnidirectional precoding for {3D} massive
  {MIMO} with uniform planar arrays,'' \emph{IEEE Trans. Wireless Commun.},
  vol.~19, no.~4, pp. 2628--2642, Apr. 2020.

\bibitem{Jiang_19}
Y.~Jiang, F.~Li, X.~Wang, and J.~Li, ``Autocorrelation complementary
  matrices,'' in \emph{Proc. 53rd Asilomar Conf. on Signals, Syst., and
  Comput.}, Nov. 2019, pp. 1596--1600.

\bibitem{Li_19}
Y.~Li, C.~Liu, R.~Zhang, Y.~Ruan, and T.~Li, ``Omnidirectional space–time
  block coding design for three-dimensional massive {MIMO} systems,''
  \emph{IEEE Access}, vol.~7, pp. 141\,701--141\,714, 2019.

\bibitem{Su2019}
D.~Su, Y.~Jiang, X.~Wang, and X.~Gao, ``Omnidirectional precoding for massive
  {MIMO} with uniform rectangular array - {Part I}: Complementary codes-based
  schemes,'' \emph{IEEE Trans. Signal Process.}, vol.~67, no.~18, pp.
  4761--4781, Sep. 2019.

\bibitem{Li2021}
F.~Li, Y.~Jiang, C.~Du, and X.~Wang, ``Construction of {Golay} complementary
  matrices and its applications to {MIMO} omnidirectional transmission,''
  \emph{IEEE Trans. Signal Process.}, vol.~69, Mar. 2021.

\bibitem{Luke_85}
H.~D. L{\"u}ke, ``Sets of one and higher dimensional {W}elti codes and
  complementary codes,'' \emph{IEEE Trans. Aerosp. Electron. Syst.}, vol.
  AES-21, no.~2, pp. 170--179, Mar. 1985.

\bibitem{Khamy2004}
S.~E. El-Khamy, M.~A. Mokhtar, and N.~O. El-Ganainy, ``New techniques for image
  change and motion detection based on complete complementary code arrays,'' in
  \emph{21st National Radio Science Conference (NRSC2004)}, Abano Terme, Italy,
  Mar. 2004, pp. 1--8.

\bibitem{2DCDMA_2004}
M.~Turcsány and P.~Farkaš, ``New {2D-MC-DS-SS-CDMA} techniques based on
  two-dimensional orthogonal complete complementary codes,'' in
  \emph{Multi-Carrier Spread-Spectrum.}, Dordrecht, The Netherlands: Springer,
  Jan. 2004, pp. 49--56.

\bibitem{2D_QCCC_CDMA}
M.~Turcsány, P.~Farkaš, P.~Duda, and J.~Kralovic, ``Performance evaluation of
  two-dimensional quasi orthogonal complete complementary codes in fading
  channels,'' in \emph{Proc.\ Joint IST Workshop on Mobile Future, Symp. Trends
  Commun. (SympoTIC)}, Bratislava, Slovakia, Jun. 2006, pp. 84--87.

\bibitem{2D_CCC}
P.~Farkaš and M.~Turcsány, ``Two-dimensional orthogonal complete
  complementary codes,'' in \emph{Proc.\ Joint 1ST Workshop on Mobile Future
  and Symp. on Trends in Commun. (SympoTIC)}, Bratislava, Slovakia, Oct. 2003,
  pp. 21--24.

\bibitem{Pai_20}
C.-Y. Pai and C.-Y. Chen, ``Constructions of two-dimensional {Golay}
  complementary array pairs based on generalized {Boolean} functions,'' in
  \emph{Proc.\ {IEEE} Int.\ Symp. Inf. Theory}, Virtual, Jun. 2020, pp.
  2931--2935.

\bibitem{Pai_22}
------, ``Two-dimensional {G}olay complementary array pairs/sets with bounded
  row and column sequence {PAPR}s,'' \emph{{IEEE} Trans. Commun.}, 2022, early
  access, DOI: 10.1109/TCOMM.2022.3166903.

\bibitem{Das_20}
S.~Das and S.~Majhi, ``Two-dimensional {Z}-complementary array code sets based
  on matrices of generating polynomials,'' \emph{{IEEE} Trans. Signal
  Process.}, vol.~68, pp. 5519--5532, 2020.

\bibitem{Wang_2021}
Z.~Wang, D.~Ma, G.~Gong, and E.~Xue, ``New construction of complementary
  sequence (or array) sets and complete complementary codes,'' \emph{{IEEE}
  Trans. Inf. Theory}, vol.~67, no.~7, pp. 4902--4928, Jul. 2021.

\bibitem{GCAP_Fiedler_08}
F.~Fiedler, J.~Jedwab, and M.~G. Parker, ``A multi-dimensional approach for the
  construction and enumeration of {Golay} complementary sequences,'' \emph{J.
  Combinat. Theory A}, vol. 115, no.~5, pp. 753--776, Jul. 2008.

\bibitem{Pai_21}
C.-Y. Pai, Y.-T. Ni, and C.-Y. Chen, ``Two-dimensional binary {Z}-complementary
  array pairs,'' \emph{{IEEE} Trans. Inf. Theory}, vol.~67, no.~6, pp.
  3892--3904, Jun. 2021.

\bibitem{Roy_20}
A.~Roy, P.~Sarkar, and S.~Majhi, ``A direct construction of $q$-ary 2-{D}
  {Z}-complementary array pair based on generalized {B}oolean functions,''
  \emph{{IEEE} Commun. Lett.}, vol.~25, no.~3, pp. 706--710, Mar. 2021.

\bibitem{Pai_212}
C.-Y. Pai and C.-Y. Chen, ``A novel construction of two-dimensional
  {Z}-complementary array pairs with large zero correlation zone,''
  \emph{{IEEE} Signal Process. Lett.}, vol.~28, May 2021.

\bibitem{Tarokh_99}
V.~Tarokh, H.~Jafarkhani, and A.~R. Calderbank, ``Space-time block codes from
  orthogonal designs,'' \emph{{IEEE} Trans. Inf. Theory}, vol.~45, no.~5, pp.
  1456--1467, Jul. 1999.

\end{thebibliography}


\end{document}